\documentclass[10 pt, conference]{ieeeconf}
\IEEEoverridecommandlockouts                              
\usepackage{graphicx}
\usepackage{amsmath,amssymb,amsfonts}
\usepackage{subcaption}
\usepackage{color}
\usepackage{placeins}
\usepackage{makecell,tabularx}

\newtheorem{definitionA}{Definition A.} 

\newtheorem{definitionB}{Definition B.} 

\usepackage{url}
\usepackage{multirow}

\usepackage{fixltx2e}
\usepackage{orcidlink}
\usepackage[outdir=./]{epstopdf}
\begin{document}

\twocolumn

\title{\LARGE \bf
Unveiling Dynamics and Patterns: A Comprehensive Analysis of Spreading Patterns and Similarities in Low-Labelled Ransomware Families
}

\author{
\authorblockN{Francesco Zola\authorrefmark{1}\orcidlink{0000-0002-1733-5515}, Mikel Gorricho\authorrefmark{1}, Jon Ander Medina\authorrefmark{1}\orcidlink{0009-0008-1107-0617}, Lander Segurola\authorrefmark{1}\orcidlink{0000-0003-4278-9081}, Raul Orduna Urrutia\authorrefmark{1}\orcidlink{0000-0002-5932-0987}}
\authorblockA{\authorrefmark{1} Vicomtech, Basque Research and Technology Alliance (BRTA)\\ Paseo Mikeletegi 57, 20009 Donostia/San Sebastian, Spain\\ \{fzola, mgorricho, jamedina, lsegurola, rorduna\}@vicomtech.org}
}

\maketitle
\thispagestyle{empty}
\pagestyle{empty}

\begin{abstract}
Ransomware has become one of the most widespread threats, primarily due to its easy deployment and the accessibility to services that enable attackers to raise and obfuscate funds. This latter aspect has been significantly enhanced with the advent of cryptocurrencies, which, by fostering decentralisation and anonymity, have transformed this threat into a large-scale outbreak. However, recent reports indicate that a small group of individuals dominate the ransomware ecosystem and try to obfuscate their activity using multiple strains characterised by a short time to live. This scenario suggests that different strains could share mechanisms in ransom collection, fund movement, and money laundering operations. For this reason, this study aims to analyse the address-transaction graphs generated in the Bitcoin network by low-labelled ransomware families. Our goals are to identify payment spreading patterns for evaluating the evolution of ransomware families and to detect similarities among different strains that potentially can be controlled by the same attacker. Specifically, this latter task assigns an address behaviour to each node in the address-transaction graphs according to its dynamics. The distribution of the behaviours in each strain is finally used to evaluate the closeness among different ransomware families. Our findings show that although ransomware families can quickly establish connections with millions of addresses, numerous families require multiple-step analysis. Furthermore, the study demonstrates that the introduced behaviours can effectively be used to highlight similarities among different ransomware strains. The outcome shows that families are similar primarily due to behaviours usually associated with ransom collection and money laundering operations. To the best of our knowledge, this work contributes to dissecting the evolution of ransomware strains and detecting distinctive markers they show within the Bitcoin network.

\end{abstract}

\IEEEoverridecommandlockouts
\begin{keywords}
Ransomware, Spread Evolution, Behaviour Distribution, Similarities, Bitcoin Analysis
\end{keywords}

\section{Introduction}
In the last two decades, ransomware has emerged as one of the most widespread threats, locking or encrypting computers and data using symmetric and asymmetric encrypting algorithms while asking for the payment of a ransom \cite{Okaneransom, richardson2017ransomware}. It operated primarily as a form of common vandalism among cybercriminals \cite{SRINIVASAN20177} until it became a sophisticated criminal practice, for example, utilising Ransomware as a Service (RaaS) models, as described by Europol in \textit{The 2023 Internet Organised Crime Threat Assessment (IOCTA)} \cite{europol2023iocta}. In fact, this report shows that the presence of cybercrime services within the network brought together and increased the number of criminal actors.

Over the years, ransomware attacks have also changed their primary targets, starting from ordinary end-users; they pass to compromise governments, banks, schools, hospitals, and business organisations in almost any sector \cite{DBLP:journals/corr/abs-2102-06249}. In the beginning, victims of ransom were directed to make payments through methods such as SMS text messages and by sending pre-paid cards via mail, calling a premium-rate telephone number, etc. \cite{richardson2017ransomware}. However, with the advent of cryptocurrencies, we witnessed the first large-scale outbreak of ransomware \cite{Okaneransom}. In fact, by promoting decentralisation and anonymity (or pseudo-anonymity in some cases), these digital assets have created an ideal environment for the ransomware business. In particular, Bitcoin and Monero are the top digital currencies used in ransomware operations \cite{evenepoel2022tracing}. On the one hand, Bitcoin represents the most known cryptocurrency, with a very high market value, and especially the easiest to acquire for victims without a tech background. On the other hand, Monero is a privacy coin that increases user and transaction anonymity, ideal for laundering activities \cite{custers}. Furthermore, the relationship between the ransomware impact, their variants, and the crypto-ecosystem has been studied and demonstrated in many researches \cite{zimba2019economic, CONTI2018162}. 

Despite the number of ransomware variants rising to over 10,000 strains (or families) in the first half of 2022, the average lifespan of each strain dropped to 70 days, indicating hidden tactics and the use of multiple strains by attackers, as suggested in \cite{chainalysis2023crypto}. Furthermore, the document indicates that a small group of individuals dominate the ransomware ecosystem. \textit{The 2024 Crypto Crime Report} \cite{chainalysis2024crypto} reports that ransomware actors linked to specific strains continuously launch new variants to make victims more willing to pay. Indeed, this ''rebranding" strategy enables ransomware attackers to distance themselves from strains that are publicly associated with sanctions or have attracted excessive scrutiny. This scenario would lead to the hypothesis that different strains controlled by a concrete attacker could share similar mechanisms for their activities, such as ransom collection, fund movement, and money laundering. If verified, this could allow Law Enforcement Officers (LEOs) to anticipate attackers' strategies, helping them track different strains towards known entities that implement Know Your Customer (KYC) policies useful for extracting real-world information, and, at the same time, could allow them to optimise their approaches by using investigation techniques followed in previous successful and similar cases.

For this reason, in this paper, we propose to study the payment spreading patterns that ransomware families generate within the Bitcoin blockchain. Our idea is to analyse and characterise the different families by evaluating their temporal evolution and, simultaneously, the dynamics generated by each address involved in the transactions. In this way, we aim to detect similarities (and differences) between ransomware variants and, so, verify if they effectively share mechanisms for their activities. Specifically, we first propose delineating distinct spreading patterns based on the population reached by each family. To achieve this, we compute the address-transaction graph \cite{fleder2015bitcoin} to represent the dynamics of the ransomware. Thus, we study its temporality, evaluating the time required for the strain to spread throughout the population. Our analysis is designed to leverage the identified spreading patterns as a lens through which we can observe and understand the evolutionary trajectories of these ransomware families. Moreover, address-transaction graphs are also used to introduce address \textit{behaviours}, i.e., labels used to identify common dynamics of the addresses within the network. The analysis demonstrates that these topological dynamics can effectively characterise each strain. Therefore, these behaviours and their distributions in each ransomware become instrumental in verifying the similarities and distinctions among the diverse families. In this work, low-labelled ransomware families, i.e., families characterised by a scant number of labelled/tagged addresses (or \textit{seeds}), are used during the analyses. The motivation for this decision and the possible limitations are discussed in Section \ref{subsec:dataset}.


\section{Background} \label{sec:related}

Traditional ransomware analysis aims to study how they are developed/programmed \cite{MORATO201814}, \cite{jung2018ransomware}. However, these attacks are continuously evolving, and hackers always try new mechanisms to deceive their victims and spread the attacks, such as leveraging new attack vectors, incorporating advanced encryption algorithms, expanding the number of file types it targets, using anonymous networks for moving the ransom, etc. \cite{kotov2023understanding}. In this sense, the success of cryptocurrencies, in part related to the pseudo-anonymity they guarantee, has converted them into the principal way to demand ransomware payments \cite{Turner2023}. This scenario has led researchers to follow the ransom within the crypto-ecosystem and study how the funds are moved, with the main goal of implementing new ransomware detection systems. This is the case of \cite{DBLP:journals/corr/abs-1906-07852}, in which an efficient data analytics framework is designed to detect new malicious addresses automatically. A ransomware detection system is presented in \cite{9527414}, employing the backpropagation artificial neural network method with Weka software. In a similar way, in \cite{electronics10172113}, a Bitcoin transaction predictive system for classifying ransomware payments has been presented, while in \cite{pham2017anomaly}, a study on detecting suspicious users and transactions using machine learning models is presented. Huang et al. \cite{8418627} introduced a measurement framework for ransomware payments, victims, and operators. In the research, they collected a variety of data sources, including ransomware binaries, seed ransom payments, victim telemetry from infections, and an extensive database of Bitcoin addresses. In \cite{turner2020discerning}, graph algorithms are used to analyse Bitcoin ransomware graphs for preventing attacks.


On the other hand, other studies are more directed at studying ransomware evolution and how they are spread within the network. This is the case of \cite{10.1145/3494557}, where authors aim to track ransom payments and the trajectory of the funds across different industries. In \cite{turner2021follow}, authors investigate how crypto exchanges and mixers are used (and combined) to launder ransomware illegal funds, while transactions associated with 35 ransomware families are further analysed in \cite{paquet2019ransomware}, using GraphSense tool for revealing the impact from an economic point of view. 

Inspired by these last works, in this paper, we aim to evaluate if different strains follow similar evolution trajectories and apply similar strategies for their activities. In particular, we analyse low-labelled ransomware families with the aim to a) identify spreading patterns based on the family's capacity to reach a large population within a fixed timeframe, b) study the temporal evolution of each family, c) define address behaviours based on the topological structures families generated in the address-transaction graph, d) demonstrate that these behaviours are enough for characterising each family, and e) finally, exploit behaviour distributions to find similarity among the different families.
 
\section{Experimental Framework}

\subsection{Dataset}\label{subsec:dataset}
\begin{figure}[!htbp]
  \centering
   \includegraphics[width=\linewidth]{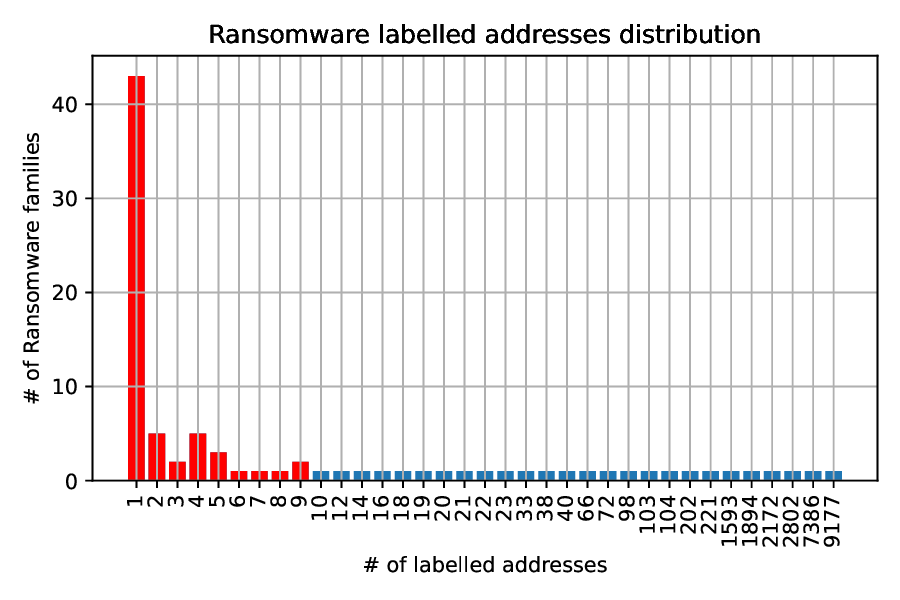}
  \caption{Seed distribution in the gathered dataset.}
  \label{fig:distribution}
\end{figure}

In this paper, the entire Bitcoin blockchain data until the block 810,000 are downloaded, i.e., all the transactions until September 30th, 2023 (about 900M transactions). On the other hand, Graphsense\footnote{https://graphsense.info/} \cite{Haslhofer:2021a}, Ransomware Payments dataset \cite{paquet2019ransomware}, and BitcoinHeist dataset \cite{DBLP:journals/corr/abs-1906-07852} have been used to collect addresses labelled as ransomware (seeds). These sources represent valid solutions used in many researches \cite{paquet2019ransomware, sun2022bitanalysis}, and allowed us to gather about 26,300 Bitcoin addresses tagged as ransomware of 89 different families, as shown in Figure \ref{fig:distribution}. 

As mentioned, this work is focused only on low-labelled families, i.e., those strains with less than 10 labelled addresses. This decision was made after considering several factors:
\begin{enumerate}
 \item families with a large number of seeds have been extensively studied in existing literature \cite{paquet2019ransomware, sun2022bitanalysis};
 \item given that RaaS involves mass production, we can infer that families containing numerous labelled addresses (seeds) can likely be interconnected due to extensive RaaS operations behind the scenes. Therefore, the results of this analysis could be skewed;
 \item as depicted in Figure \ref{fig:distribution}, families with more than 10 seeds are not the most common ones;
 \item low-labelled families may have a limited number of seeds for several reasons. They could belong to smaller ransomware operations that are not prominent within the industry, or they might not have been thoroughly discovered. Additionally, they could employ sophisticated mechanisms that make them difficult to detect, thereby increasing their value.
\end{enumerate}

The constraint left us with a dataset of 137 addresses of 65 different families (Table \ref{tab:familiydistribution}). Although the small number of addresses could seem a limitation in the validation of the results, it is to be noted that the dataset still contains $\sim$72\% of the original ransomware families. This aspect is the most relevant one since the main goal of this study is to check and evaluate similarities among the spreading patterns and the behaviour distribution of different ransomware families.
 similarities among different families.

Once the dataset is reduced, all the seeds are used as a starting point for extracting each ransomware's address-transaction graph (Section \ref{addressgraph}). In particular, one single graph is extracted for each family, so in the cases of multiple seeds, the generated graphs are merged into a single one.

\begin{table}
\resizebox{0.9\linewidth}{!}{
\centering
\begin{tabularx}{\linewidth}{lcX}
\hline
\textbf{Year} &\textbf{\# Seeds} & \textbf{Ransomware family name}\\
\hline
\multirow{1}{*}{2015} &1 & \textit{chimera, teslacrypt}\\
\hline
\multirow{5}{*}{2016} &1 &\textit{7ev3n, bucbi, comradecircle, cryptohost, ctb-locker, ecovector, exotic, nullbyte, phoenix, popcorntime, sam}\\
&4 &\textit{cryptohitman}\\
&5 & \textit{apt, xtplocker} \\
&7 & \textit{towerweb} \\
& 8 & \textit{cryptconsole,venuslocker} \\
\hline
\multirow{4}{*}{2017} & 1 &\textit{black mamba, globeimposter, lamdalocker, lockon, ransomnix, spora, storagecrypter, vevolocker, wannasmile, xlocker}\\
&3 &\textit{xorist} \\
& 4 &\textit{xlockerv5.0} \\
& 6 & \textit{hc6/hc7}\\
\hline
\multirow{1}{*}{2018} & 1 &\textit{black ruby, blackrouter, gula, qweuirtksd} \\
\hline
\multirow{2}{*}{2019} &1 &\textit{encrpt3d, git, tejodes}\\
&2 &\textit{decryptiomega} \\
\hline
\multirow{6}{*}{2020} & 1 & \textit{ako, bitpaymer, kelly, mountlocker, wannaren}\\
& 2 & \textit{black kingdom}\\
& 3  &\textit{lockbit}\\
& 4 &\textit{ragnarlocker}\\
& 7 & \textit{rEvil}\\
&9 & \textit{egregor} \\
\hline
\multirow{2}{*}{2021} & 1 &\textit{albdecryptor, avaddon, avoslocker, bagli, blackmatter, chupacabra, hellokitty, ranzy locker, suncrypt, vega}\\
&3 &\textit{darkside} \\
\hline
\multirow{2}{*}{2022} &1 & \textit{darkangels, hive, quantum}\\
&9 & \textit{deadbolt}\\
\hline
 &\textbf{137} &\textbf{65 families} \\
\hline
\end{tabularx}}
\caption{Ransomware families used in the analysis separated per year and number of seeds.}
\label{tab:familiydistribution}
\end{table}

\subsection{Address-Transaction graph} \label{addressgraph}
Cryptocurrencies have an inherent structure of information that allows it to be represented in a directed graph. In fact, since transactions are used for connecting two or more crypto addresses, an address-transaction graph can be implied \cite{fleder2015bitcoin}. This graph approximates the cryptocurrency flow by connecting public key addresses over time. The vertices within the graph represent addresses and transactions. Thus, directed edges (arrows) between addresses and transactions represent incoming relations, while directed edges between transactions and addresses denote outgoing relations. Each edge may incorporate additional features like amount, fee, timestamps, etc. 


In this research, we create specific address-transaction graphs starting from the seeds and exploring their neighbours in $n$ steps (or hops). Let $X1$ be a seed to be used as a starting point; its $n$-step address-transaction graph is a graph in which all the paths from $X1$ involve maximum $n$ transactions; therefore, they have a maximum length of $2n$. These paths must include $X1$ as a starting/ending point, and consequently, they should end/start with an address node. 

\section{Ransomware Payment Spreading Pattern}

To investigate how ransomware evolves and is spread within the Bitcoin network, 2-step address-transaction graphs have been analysed. Through this task, we have identified and utilised four spreading patterns: \textit{slow, moderate, fast} and \textit{extremely fast (or exFast)}. The first category includes families that reach less than 500 distinct addresses in the 2-step graphs; the second includes families that link from 500 to 50,000 addresses; the third one connects 50,000 to 500,000 addresses; and the last one reaches more than 500,000 addresses. 

\begin{figure*}[!htbp]
  \centering
  \begin{subfigure}[b]{0.42\linewidth}
   \includegraphics[width=\linewidth]{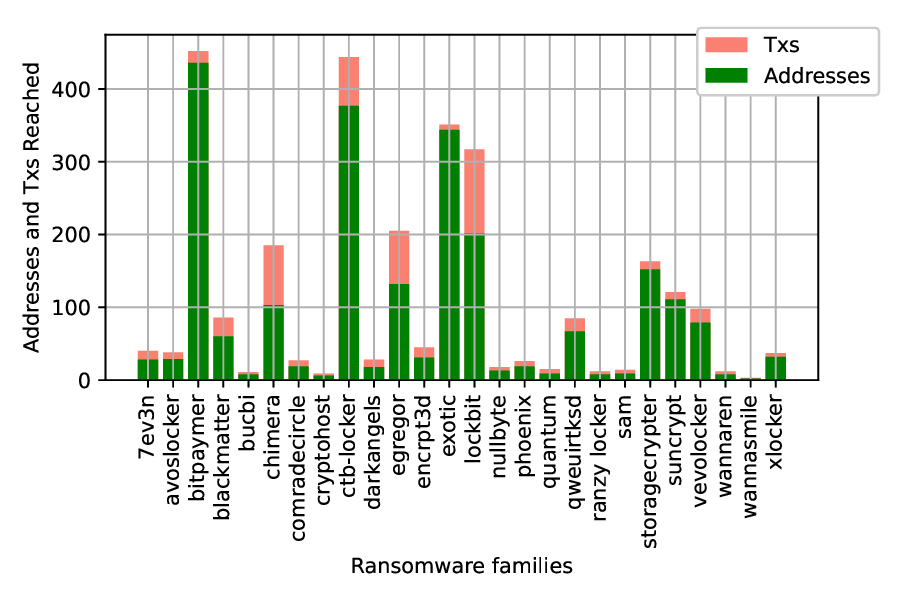}
    \caption{Slow pattern ($<$500)}
    \label{fig:slow}
  \end{subfigure}
  \begin{subfigure}[b]{0.42\linewidth}
    \includegraphics[width=\linewidth]{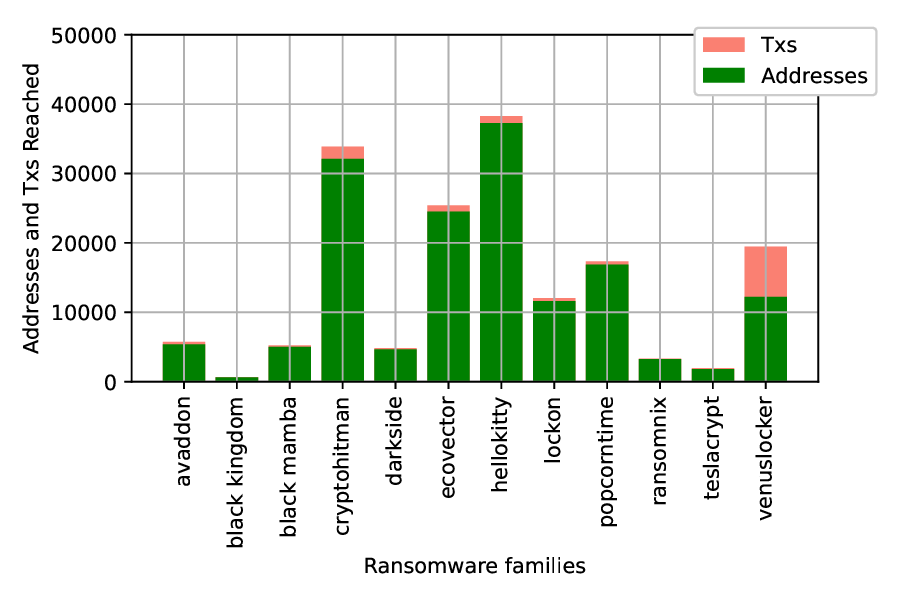}
    \caption{Moderate pattern (500-50,000)}
    \label{fig:moderate}
  \end{subfigure}
  \begin{subfigure}[b]{0.42\linewidth}
    \includegraphics[width=\linewidth]{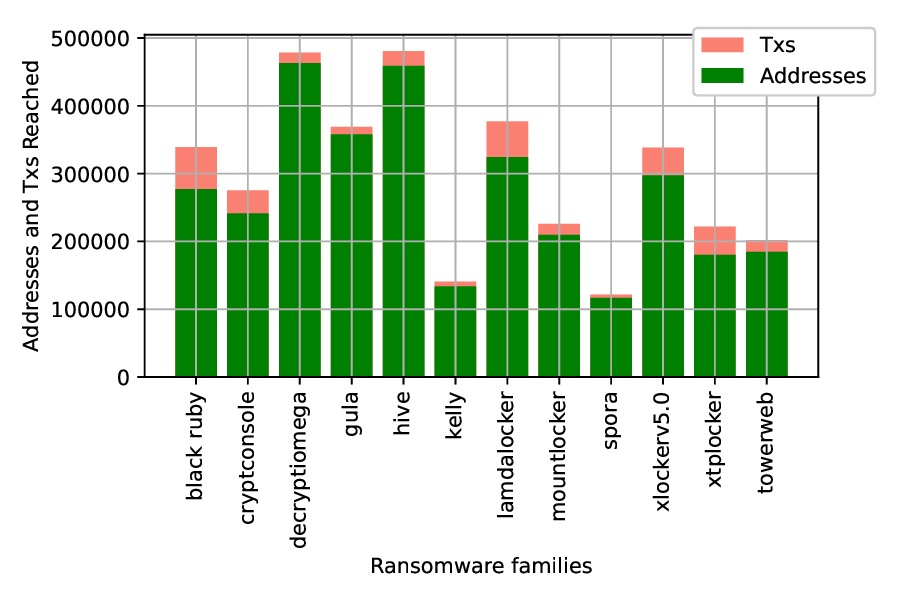}
    \caption{Fast pattern (50,000-500,000)}
    \label{fig:fast}
  \end{subfigure}
  \begin{subfigure}[b]{0.42\linewidth}
   \includegraphics[width=\linewidth]{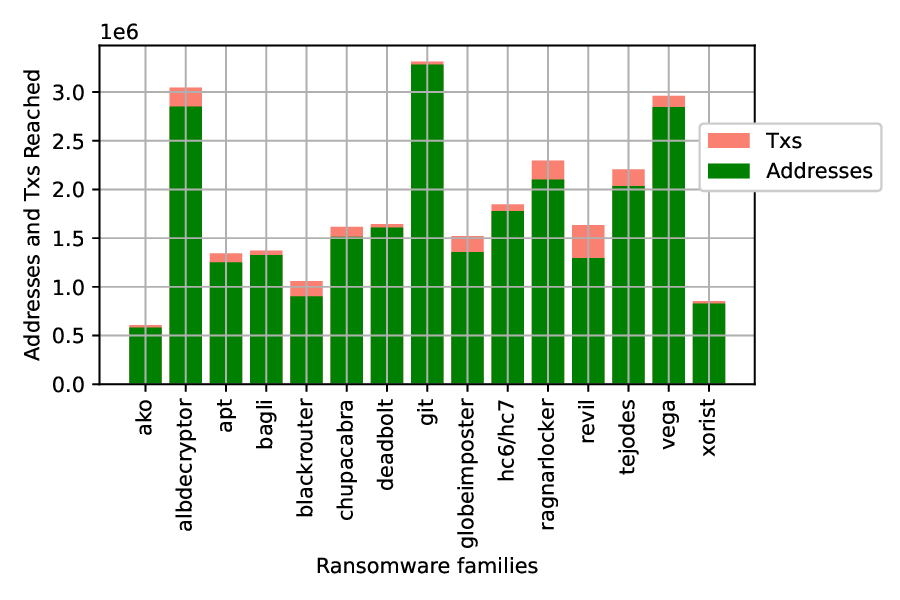}
    \caption{Extremely Fast pattern ($>$500,000)}
    \label{fig:exfast}
  \end{subfigure}
  \caption{Ransomware payment spreading patterns.}
  \label{fig:spreadinggraph}
\end{figure*}

Figure \ref{fig:slow} shows that the majority of the families - 26 families out of the 65 considered ($40$\%) - follow the \textit{slow} pattern in the 2-step graphs, while 12 families ($\sim19$\%) use a \textit{moderate} spreading pattern (Figure \ref{fig:moderate}). A similar population ($\sim19$\%) is shown in Figure \ref{fig:fast}, where families with a \textit{fast} spreading pattern are reported. Finally, Figure \ref{fig:exfast} shows that $23$\% of the considered ransomware families (15 out of 65) have an \textit{exFast} spreading pattern. These strains involve more than 500,000 addresses with peaks of $\sim$3M, as is the case of \textit{albdecryptor, git} and \textit{vega} families. 


\begin{figure*}[!htbp]
  \centering
  \begin{subfigure}[b]{0.42\linewidth}
   \includegraphics[width=\linewidth]{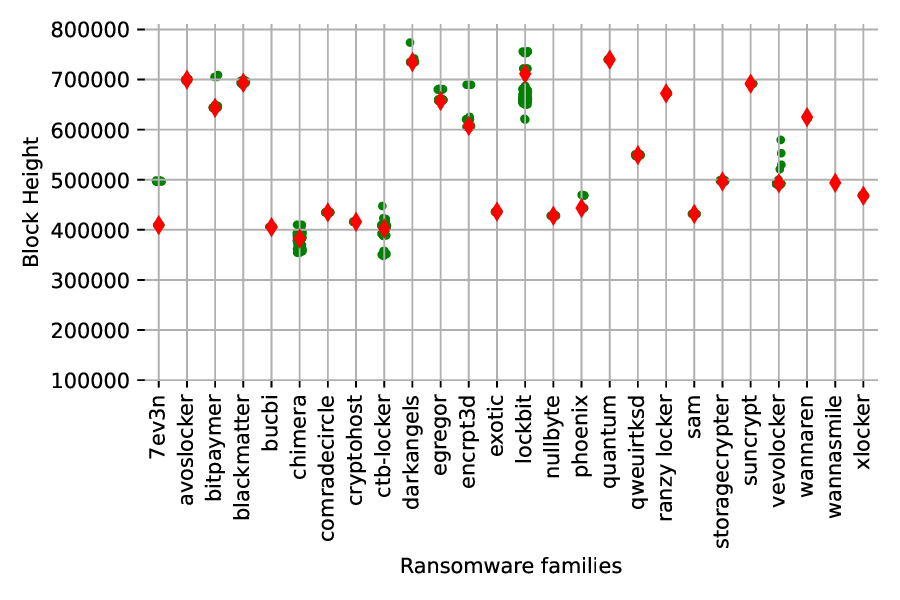}
    \caption{Families with slow pattern}
    \label{fig:slow_block}
  \end{subfigure}
  \begin{subfigure}[b]{0.42\linewidth}
    \includegraphics[width=\linewidth]{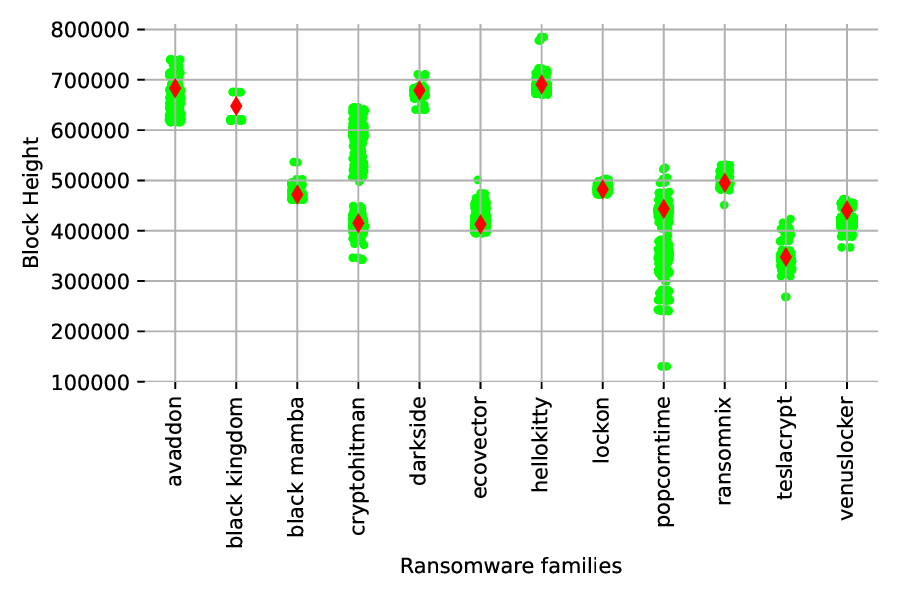}
    \caption{Families with moderate pattern}
    \label{fig:moderate_block}
  \end{subfigure}
  \begin{subfigure}[b]{0.42\linewidth}
    \includegraphics[width=\linewidth]{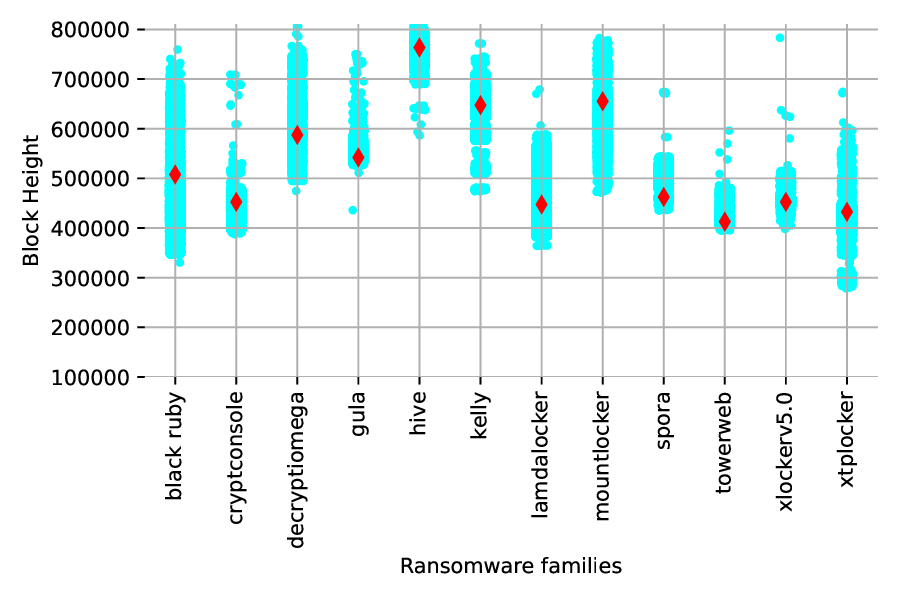}
    \caption{Families with fast pattern}
    \label{fig:fast_block}
  \end{subfigure}
  \begin{subfigure}[b]{0.42\linewidth}
   \includegraphics[width=\linewidth]{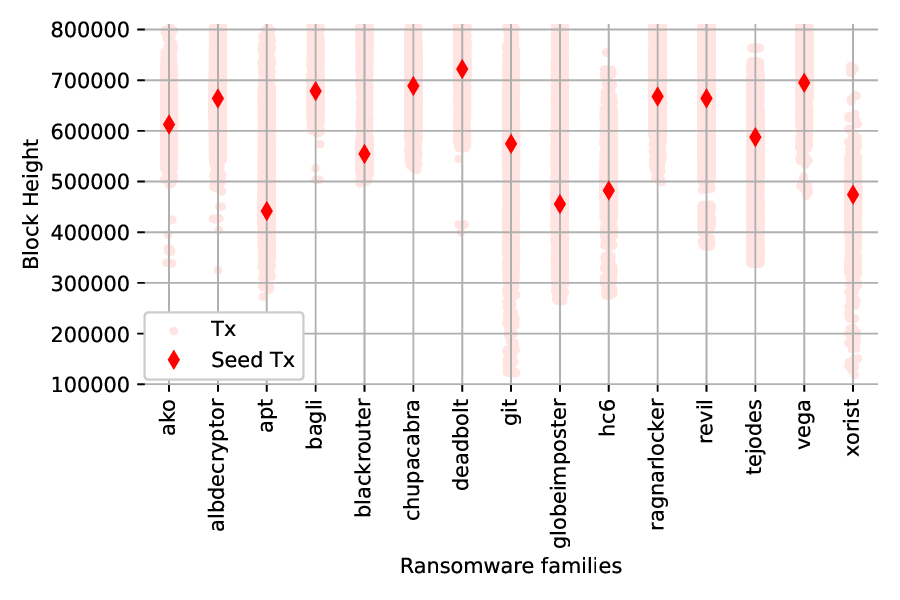}
    \caption{Families with exFast pattern}
    \label{fig:exfast_block}
  \end{subfigure}
  \caption{Block height of the transactions within the 2-step graphs for each ransomware family. Highlighted the transactions where the seeds are involved.}
  \label{fig:spreadinggraph_block}
\end{figure*}

Figure \ref{fig:spreadinggraph_block} reports the block heights of the transactions discovered in the 2-step address-transaction graph of each family. This analysis allows us to highlight the temporality of the transactions related to seed addresses with the spreading pattern of the family. In general, the \textit{slow} and \textit{moderate} spreading patterns are composed mainly of transactions achieved in close proximity to the seed transaction, i.e., the transaction in which the seed appears for the first time (Figure \ref{fig:slow_block} and Figure \ref{fig:moderate_block}). On the other hand, as shown in Figure \ref{fig:fast_block}, \textit{fast}-families show two different situations: 4 strains still achieve transactions just near to the seed transaction, while the other 8 are characterised by transactions achieved over 5/7 years, i.e., their activities are spread over 200,000/300,000 blocks homogenously. Finally, Figure \ref{fig:exfast_block} shows that all the \textit{exFast}-families involve transactions continuously in time. The clearest instance is presented by the \textit{git} strain, which is able to link transactions from about block 100,000 (2010) until block 800,000 (2024), considering just the 2-step graph. Furthermore, it is interesting to highlight that 2-step graphs are enough for determining that some families, such as \textit{ecovector, ransomnix, gula, spora, towerweb, blackrouter}, mainly comprise transactions achieved \textit{after} the seed transaction, suggesting that the seed is directly connected to output operations, like the ones related to moving the funds and money laundering. On the other hand, some families also show multiple transactions in the 2-step graphs previously to the seed transaction, hinting at the presence of strong mechanisms for raising funds and collecting ransom.

\section{Graph Behavioural Analysis}

\subsection{Behaviour Similarity}

\begin{figure}[!htbp]
\centering
\includegraphics[width=\linewidth]{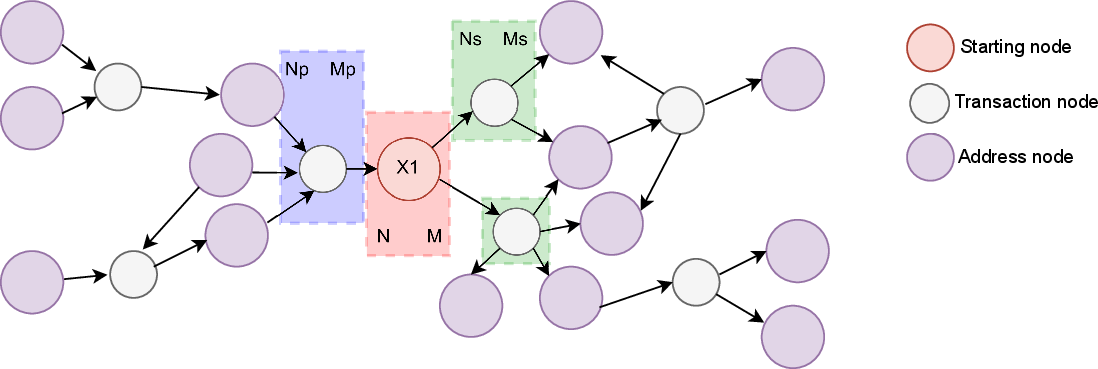}
\caption{Schema of a $n$-step address-transaction graph.}
\label{fig:schema}
\end{figure}

As mentioned in Section \ref{addressgraph}, address-transaction graphs capture all the dynamics a seed generates within the network, linking address and transaction nodes. For this analysis, the topological structure (or subgraph) of each address node is evaluated to assign a behaviour to it (transaction nodes remain unlabelled). Before introducing the available behaviours, let $X1$ be the address to be analysed, $N$ its input degree (i.e., the number of input edges coming from transaction nodes) and $M$ the output degree (i.e., the output edges to transaction nodes), depicted in Figure \ref{fig:schema}. In the same way, it is possible to define $Np$ as the input degree of a predecessor node  (transaction node) of $X1$, while $Mp$ is the output degree of the predecessor node. Finally, $Ns$ and $Ms$ are the input and output degrees of the successor node (transaction node), respectively.

\begin{figure}[!htbp]
  \centering
   \includegraphics[width=\linewidth]{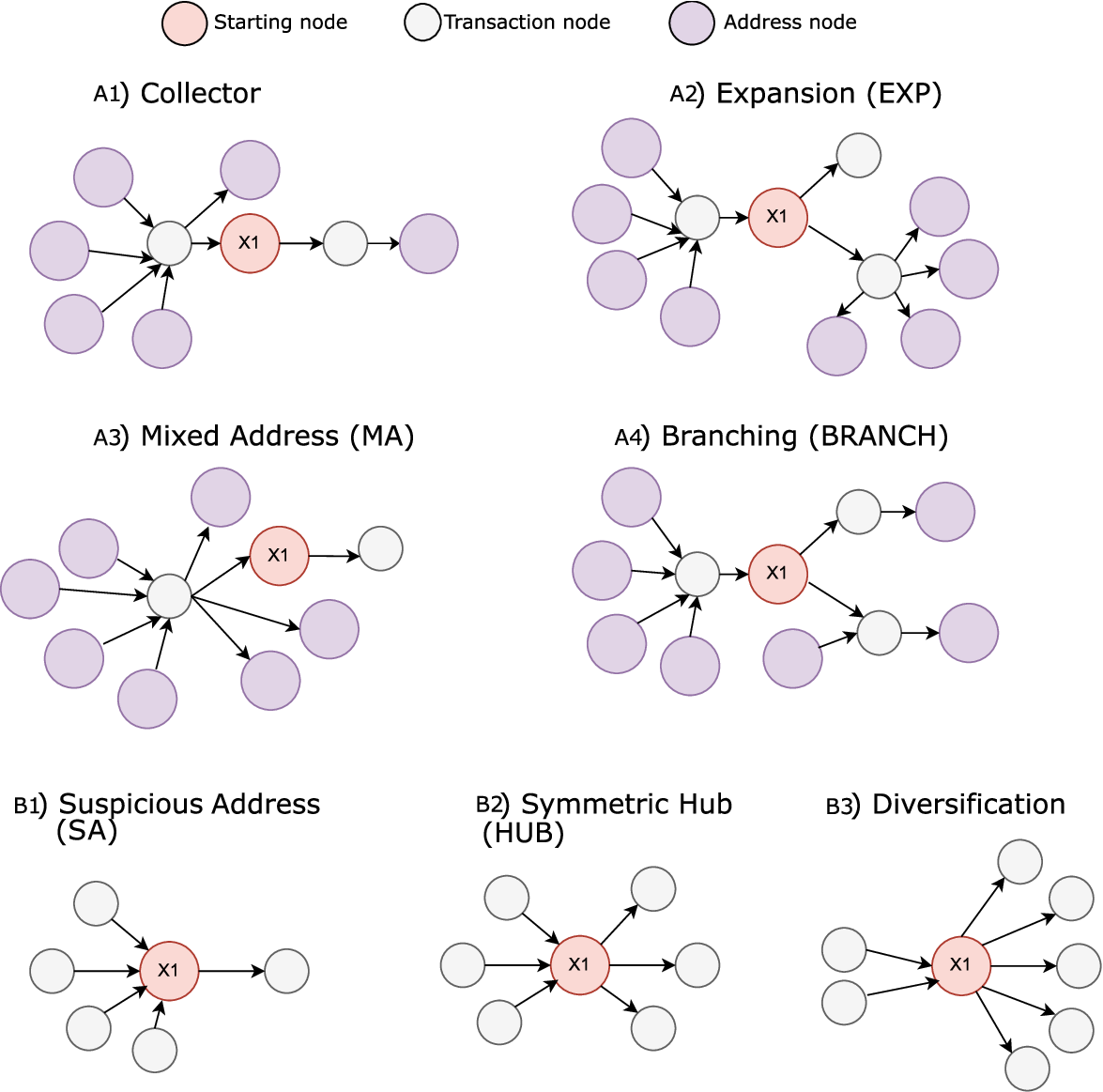}
  \caption{Euclidean distance between ransomware families distributions.}
    \label{fig:behaviours}
\end{figure}

In this paper, 7 behaviours have been defined, based on the most common topological structures. The first four (A.1, A.2, A.3, and A.4) are defined by evaluating the number of input and output addresses of the transaction nodes directly linked to the starting point ($Ns$, $Ms$, $Np$, $Mp$), while the other three behaviours (B.1, B.2, and B.3)  are defined by checking only the ratio between input and output transaction nodes linked ($N$, $M$). More specifically:

\begin{definitionA}[Collector]
        Let $X1$ be a starting address node; $X1$ is said to be a \textit{collector} node if it receives money from at least one node transaction (i.e., $N>0$) that in turn receives money from multiple addresses ($Np>3$), while $X1$ sends money to no more than one transaction ($M<2$) that in turn sends money to no more than 2 addresses ($Ms<3$). As shown in Figure \ref{fig:behaviours}-A1, this behaviour is associated with topological structures that aggregate address flows usually generated (but not only) by ransom collecting operations. 
    \end{definitionA}

    \begin{definitionA}[Expansion or EXP]
    Let $X1$ be a starting address node; it represents a \textit{EXP} if it sends money at least to one transaction ($M>0$), which in turn sends money to multiple addresses ($Ms>3$), and at the same time, $X1$ receives money from at least one node transaction ($N>0$) which in turn receive money from multiple addresses ($Np>3$). This behaviour is related to topological structures where the central address ($X1$) serves as an intermediary between multiple addresses (Figure \ref{fig:behaviours}-A2), revealing potential "money mule" situations.
    \end{definitionA}

    \begin{definitionA}[Mixed Address or MA]
    Let $X1$ be a starting address node; it represents a \textit{MA} if it sends money to no more than one transaction ($M<2$), and it receives money from at least one transaction nodes that in turn receive money from multiple addresses ($Np>3$), and, at the same time, this transaction nodes sends money not only to $X1$ but to multiple addresses ($Mp>3$). As shown in Figure \ref{fig:behaviours}-A3, this behaviour is associated with topological structures that blend address flows, potentially to obfuscate the origin of the funds. Such structures can suggest, among other possibilities, involvement in money laundering activities.
    \end{definitionA}

    \begin{definitionA}[Branching or BRANCH]
Let $X1$ be a starting address node; it represents a \textit{BRANCH} if it sends money at least to one transaction ($M>0$), which in turn sends money to no more than 2 addresses ($Ms<3$), and at the same time, $X1$ receives money from at least one transaction node ($N>0$) which in turn receive money from multiple addresses ($Np>3$). This behaviour is related to topological structures where the central address ($X1$) serves as an intermediary to sort and accumulate funds (Figure \ref{fig:behaviours}-A4). These structures can indicate potential money laundering operations.
    \end{definitionA}

        \begin{definitionB}[Suspicious Address or SA]
Let $X1$ be a starting address node; it represents a \textit{SA} if it receives money from at least 1 transaction ($N>0$) and sends money to few transactions, keeping a ratio of $\frac{M}{N}<0.5$. This condition generates topological structures, such as the one depicted in Figure \ref{fig:behaviours}-B1, where the central address ($X1$) may function as an undetected seed or as an address involved in ransom collection operations, potentially colluding with the attacker.

    \end{definitionB}

        \begin{definitionB}[Symmetric Hub or HUB]
Let $X1$ be a starting address node; it represents a \textit{HUB} if it receives money from at least 1 transaction ($N>0$) and sends money to a number of transactions, keeping a ratio of $0.5<\frac{M}{N}<1.5$. As shown in Figure \ref{fig:behaviours}-B2, this behaviour is associated with topological structures that blend multiple transaction flows, no addresses as happen for \textit{MA}. Yet, these structures can be a hint of operations aimed at obfuscating and transferring ransom funds.
    \end{definitionB}

        \begin{definitionB}[Diversification]
Let $X1$ be a starting address node; it represents a \textit{Diversification} if it receives money from at least 1 transaction ($N>0$) and sends money to a large number of transactions, keeping a ratio of $\frac{M}{N}>1.5$. As shown in Figure \ref{fig:behaviours}-B3, this behaviour is associated with topological structures that spread funds through multiple transactions. These structures may resemble, among others, cash-back strategies used to pay various involved attackers or to increase the complexity of fund traceability.
    \end{definitionB}
    
All the address nodes that do not fit any previously listed behaviours are not classified, i.e., labelled as \textit{None}. It is to be noted that each address node can show none, one or two aforementioned behaviours. In this latter case, one will be from the A-definition and the other from the B-definition.

The analysis starts by identifying the composition of each ransomware family and counting how many address nodes it has labelled for each of the previously defined behaviours. Therefore, these values are rescaled by the total number of the address nodes in the graph, including those not classified but excluding the transaction nodes. In this way, each ransomware family has been characterised by a 7-dimensional vector, where each element represents the probability that a node in the graph belongs to each behaviour. These vectors are firstly analysed in a 2-dimensional space using the Principal Component Analysis (or PCA \cite{mackiewicz1993principal}). Then, in order to evaluate the proximity and, therefore, the similarity among the different families, their distance is computed using the Euclidean distance \cite{elmore2001euclidean}. In particular, let \( \mathbf{p} = (p_1, p_2, \ldots, p_7) \) and \( \mathbf{q} = (q_1, q_2, \ldots, q_7) \) the behaviour distribution of two families, their distance is computed applying the Equation \ref{eq:1}.

\begin{figure}[!htbp]
  \centering
    \includegraphics[width=0.9\linewidth]{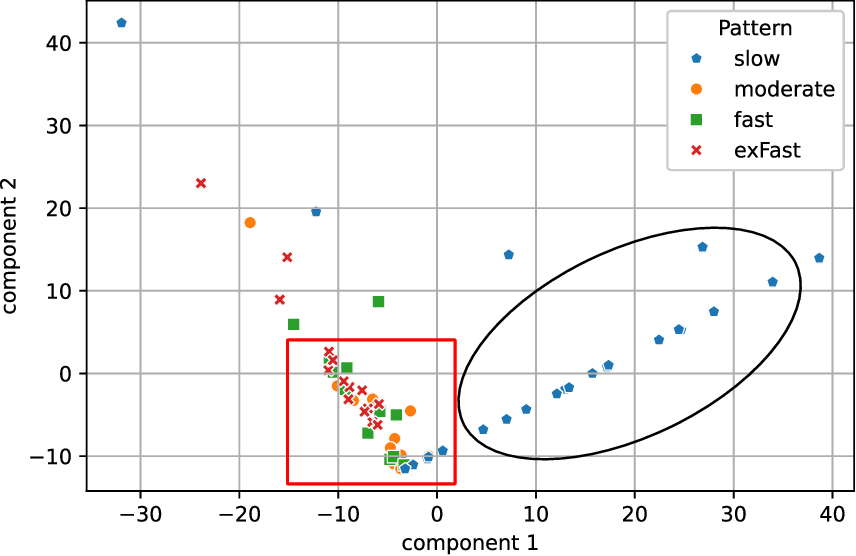}
    \caption{Visualization of all families using 2D PCA.}
  \label{fig:pca}
\end{figure}

\begin{figure*}[!htbp]
  \centering
   \includegraphics[width=\linewidth]{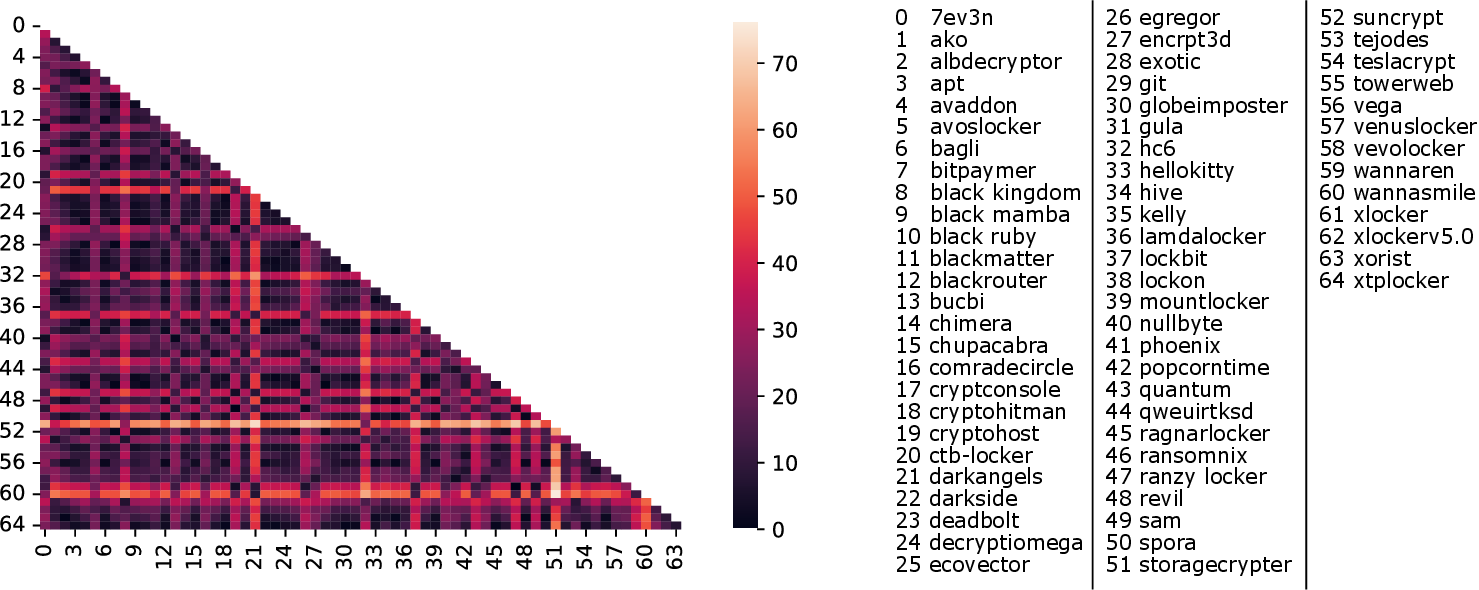}
  \caption{Euclidean distance between ransomware families distributions.}
    \label{fig:hist}
\end{figure*}

Figure \ref{fig:pca} shows the PCA of the vectors. This figure highlights that \textit{slow} elements are rare (almost linear) distribution in the space (black circle), while all the other samples are concentrated in a specific area of the space (red square). The closeness among the family distributions can be further appreciated in Figure \ref{fig:hist}, where a heatmap with the computed Euclidean distances is reported. The figure reveals that different families have a very close distribution (dark points), such as \textit{7eve3n} with \textit{bucbu}, \textit{comradecircle}, and other families. On the other hand, some strains possess a unique distribution that results in significant distances (white points), as is the case of \textit{darkangles} or \textit{storagecrypter}.

\begin{equation} \label{eq:1}
d = \sqrt{\sum_{i=1}^{7} (q_i - p_i)^2}
\end{equation}

\begin{figure*}[!htbp]
  \centering
  \begin{subfigure}[b]{0.32\linewidth}
   \includegraphics[width=\linewidth]{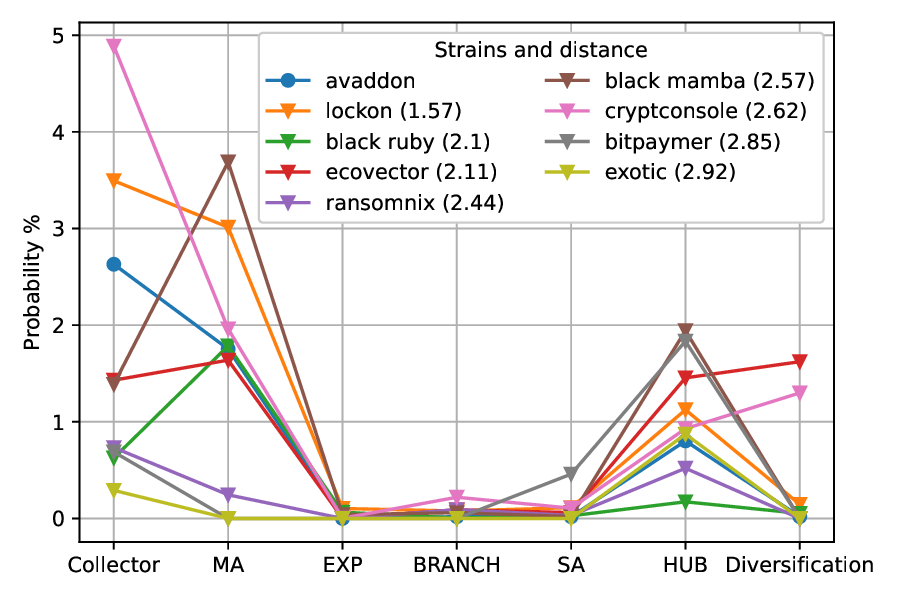}
    \caption{\textit{Avaddon} and random samples from close related families ($\lambda\leq$3.81)}
    \label{fig:prob1}
  \end{subfigure}
  \begin{subfigure}[b]{0.32\linewidth}
    \includegraphics[width=\linewidth]{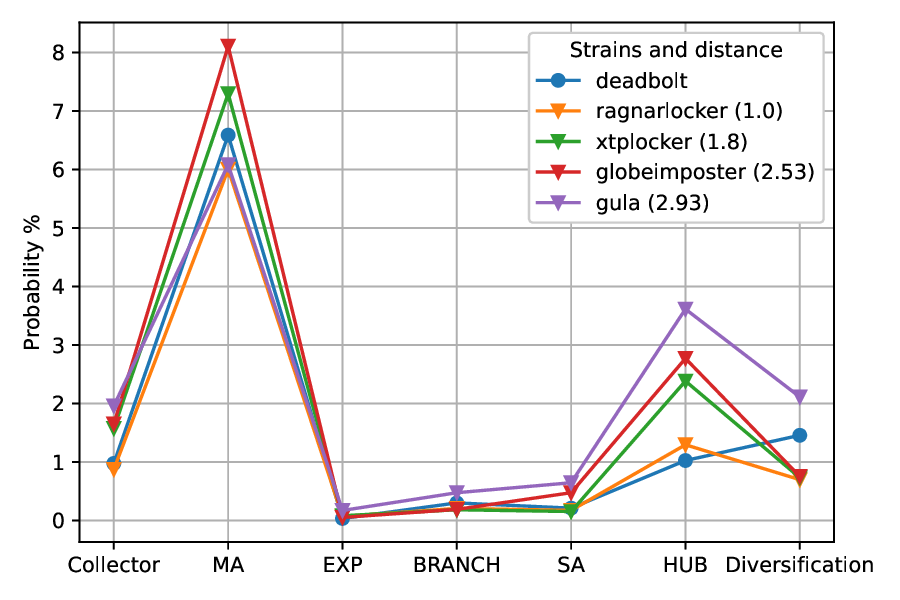}
    \caption{\textit{Deadbolt} and random samples from close related families ($\lambda\leq$3.81)}
    \label{fig:prob2}
  \end{subfigure}
  \begin{subfigure}[b]{0.32\linewidth}
    \includegraphics[width=\linewidth]{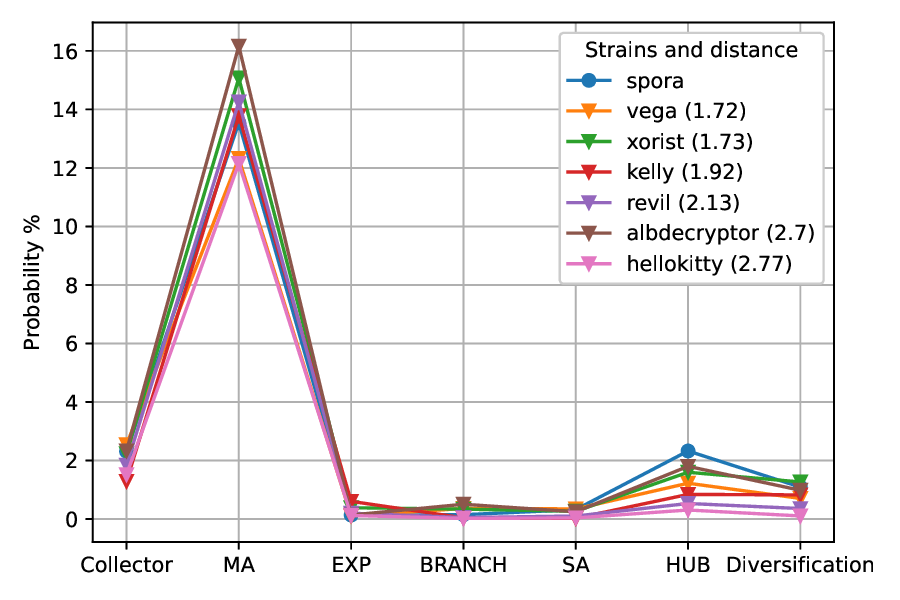}
    \caption{\textit{Spora} and random samples from close related families ($\lambda\leq$3.81)}
    \label{fig:prob3}
  \end{subfigure}
  \begin{subfigure}[b]{0.32\linewidth}
   \includegraphics[width=\linewidth]{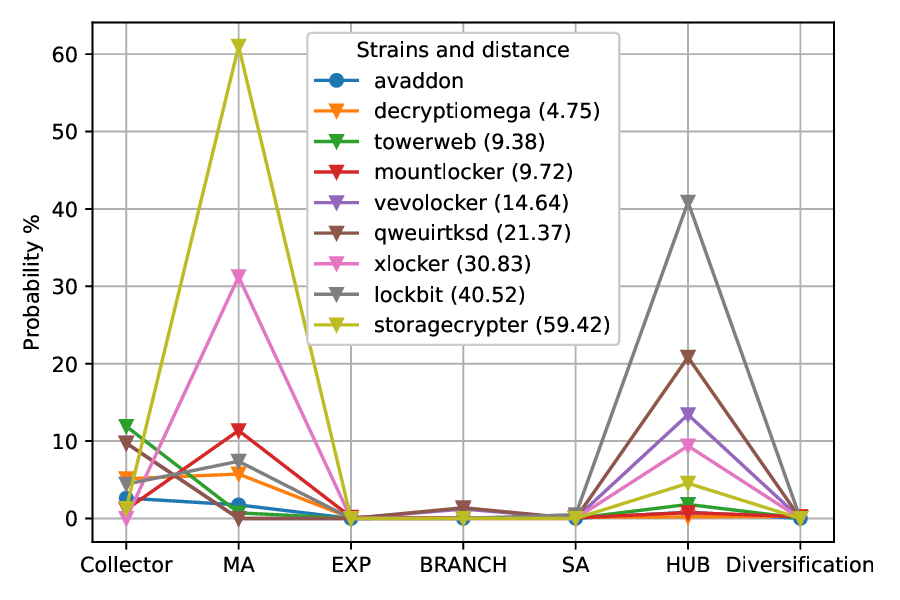}
    \caption{\textit{Avaddon} and random samples from the distant related families ($\lambda\geq$3.81)}
    \label{fig:prob11}
  \end{subfigure}
  \begin{subfigure}[b]{0.32\linewidth}
    \includegraphics[width=\linewidth]{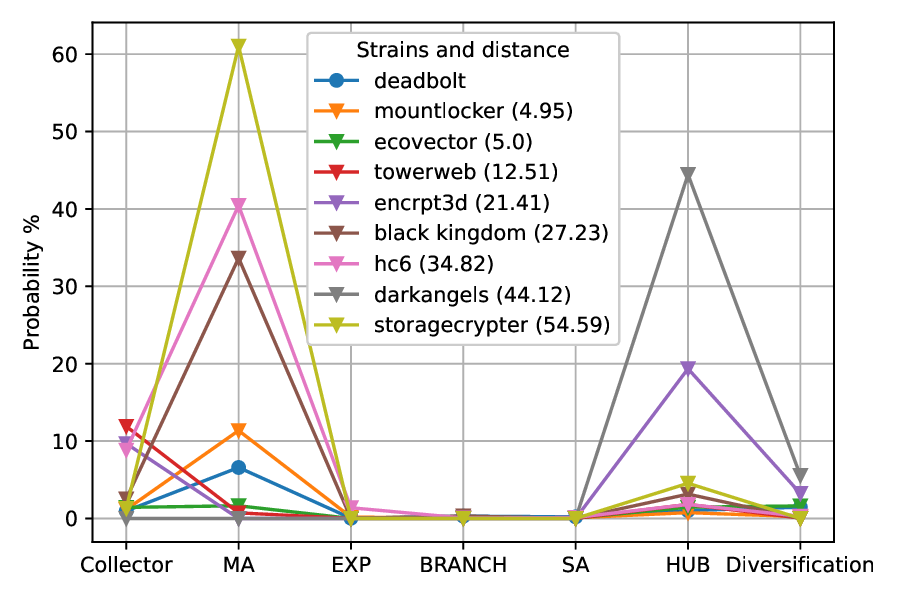}
    \caption{\textit{Deadbolt} and random samples from the distant related families ($\lambda\geq$3.81)}
    \label{fig:prob22}
  \end{subfigure}
  \begin{subfigure}[b]{0.32\linewidth}
    \includegraphics[width=\linewidth]{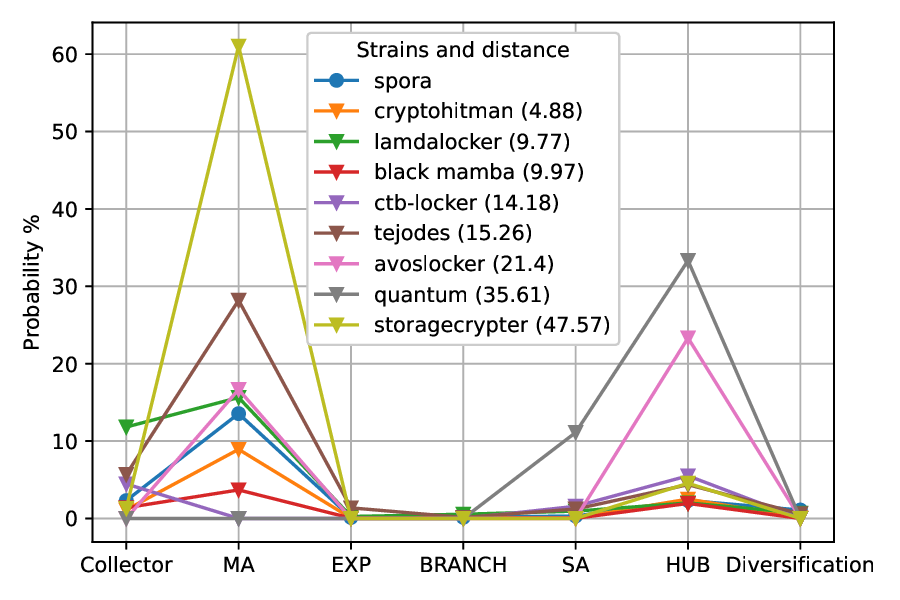}
    \caption{\textit{Spora} and random samples from the distant related families ($\lambda\geq$3.81)}
    \label{fig:prob33}
  \end{subfigure}
  \caption{Three ransomware behaviour distributions (\textit{avaddon, deadbolt,} and \textit{spora}), and their close and distant strains (sampled).}
  \label{fig:probdistance}
\end{figure*}

The distance heatmap reported in Figure \ref{fig:hist} can be used to assess similar distributions by establishing a predefined threshold ($\lambda$) and subsequently determining the number of distances that are less than/equal to $\lambda$. Specifically, the families with distances $\leq \lambda$ can be grouped in \textit{clusters}, while the others represent \textit{isolated} strains. In this analysis, values of $\lambda$ ranging from 1\% to 10\% of the maximum distance reported in the heatmap ($d_{\text{max}}$, which is 76.12) are used, as shown in Table \ref{tab:dst}. The results show that the choice of $\lambda$ affects the clusters as expected. Increasing the $\lambda$ values results in more clusters being created, with higher average populations and fewer isolated strains. This analysis confirms the closeness in the family behavioural distribution, indeed, considering only 10\% of the maximum distance, 57 families are grouped in 51 different clusters, leaving only 8 strains isolated. Yet, a validation of the closeness between ransomware behavioural distributions is reported in Figure \ref{fig:probdistance}, where three behaviour distributions of random strains are chosen as baseline (\textit{avaddon, deadbolt,} and \textit{spora}), and for each of them samples with different distances are depicted.

\begin{table*}[!htbp]\centering
\begin{tabular}{lc|cccc|cc}
\hline
\multicolumn{1}{c}{$\lambda_\%$} &\multicolumn{1}{c|}{$\lambda =\frac{\lambda_\%*d_{max}}{100}$} & \multicolumn{1}{c}{\textbf{\begin{tabular}[c]{@{}c@{}}Clusters\\number\end{tabular}}} & \multicolumn{1}{c}{\textbf{\begin{tabular}[c]{@{}c@{}}Clusters\\avg. population\end{tabular}}} & \multicolumn{1}{c}{\textbf{\begin{tabular}[c]{@{}c@{}}Clusters\\max. population\end{tabular}}} & \multicolumn{1}{c}{\textbf{\begin{tabular}[c]{@{}c@{}}Clusters\\min. population\end{tabular}}} & \multicolumn{1}{|c}{\textbf{\begin{tabular}[c]{@{}c@{}}Clustered\\strains\end{tabular}}} & \multicolumn{1}{c}{\textbf{\begin{tabular}[c]{@{}c@{}}Isolated\\strains\end{tabular}}} \\ \hline
 1 &0.76 & 3 & 2.00 & 2 & 2 &6 & 59 \\ 
 2 &1.52 & 10 & 2.20 & 3 & 2 &18 &47 \\ 
 3 &2.28 & 24 & 3.58 & 6 & 2 &21 &34 \\ 
 4 &3.05 & 30 & 5.50 & 9 & 2 &38 &27 \\ 
 5 &3.81 & 39 & 6.28 & 12 & 2 &49 &16 \\ 
 10 &7.61 & 51 & 15.24 & 29 & 2 &57 &8 \\  \hline
\end{tabular}
\caption{Distance threshold $\lambda$ with $d_{\text{max}}$=76.12.}
\label{tab:dst}
\end{table*}

Figures \ref{fig:prob2} and \ref{fig:prob3} confirm that strains with close distance (small $\lambda$) indeed exhibit similar distributions, while Figure \ref{fig:prob1} shows that, in this case, families display varying but limited variations in the probabilities of behaviours such as \textit{Collector, MA,} and \textit{HUB}. On the other hand, Figures \ref{fig:prob11}, \ref{fig:prob22} and \ref{fig:prob33} demonstrate that significant variations in these three behaviours, and in one case, also in \textit{SA}, leading to increased distance between the strains and promoting their differentiation. These results confirm once again the outcome of the previous temporal analysis since the topological structure associated with \textit{MA} behaviours is typical of mixer transactions that usually are used for money laundering operations, while \textit{SA} and \textit{Collector} structures recall mechanisms for gathering ransom from multiple victims.

\section{Conclusion}

This study aims to evaluate if different ransomware strains follow similar evolution trajectories and share similar strategies for their activities (gathering the ransom, money laundering, moving the funds, etc.). To do that, we first study the payment spreading pattern that low-labelled ransomware families generate within the Bitcoin network. The temporal analysis revealed that several families are able to quickly link millions of Bitcoin addresses, analysing just their connections in 2 steps. Indeed, they have continuously generated payment operations and activities in the Bitcoin network for years. Furthermore, we introduce $7$ specific address behaviours related to graph topologies that are used to validate commonalities that may exist among families. The study demonstrates that the introduced behaviours enable not only the characterisation of ransomware families that show quick spreading patterns but also the slower families (the ones that require more-step analyses). The study aids in detecting \textit{close} families and confirms that different strains employ comparable approaches in their operations, revealing that a crucial factor lies in the topological structure created by specific behaviours usually related to money laundering and ransom collections. 

This work represents a preliminary step, offering an intriguing yet partial understanding of the ransomware ecosystem. In fact, there is no definitive evidence to suggest that the same attacker or group is responsible for multiple low-labeled ransomware families. Additionally, high-represented families should be incorporated into the analysis to refine the approach further. Nonetheless, the results are promising, providing valuable insights into the distinctive traits that differentiate each strain in terms of temporal evolution and behaviour distribution.

As future work, it will be useful to involve topological structures created starting from other known (no-ransomware-related) entities, such as Exchanges and Mixers, in order to a) underline the validity of the behaviours defined in this work and b) study how these entities act in the ransomware propagation. On the other hand, ML-based clustering algorithms can be used over the extracted characterisation of strains to gather deeper (dis)similarity insights. These extensions will allow LEOs to develop effective ransomware tracking mechanisms and optimise their \textit{modus operandi}.

\section*{ACKNOWLEDGMENT}
This work has been partially supported by the European Union's Horizon 2020 Research and Innovation Program under the project FALCON (Grant Agreement No. 101121281)

\bibliographystyle{splncs04}
\bibliography{main}

\end{document}